\documentclass[prl,twocolumn,aps,amsfonts,showpacs,longbibliography,notitlepage,superscriptaddress]{revtex4-2}
\usepackage{graphicx}
\usepackage[dvipsnames]{xcolor}
\usepackage{rotating}
\usepackage{amsmath,amssymb,graphics,amsthm}
\usepackage{amsfonts,dsfont,mathtools}
\usepackage{bbm}
\usepackage{bm}
\usepackage{array}
\usepackage{appendix}
\newcolumntype{P}[1]{>{\centering\arraybackslash}p{#1}}
\usepackage{multirow}
\usepackage{physics}
\usepackage{tikz-cd}
\usepackage{adjustbox}
\usepackage{stmaryrd}
\usepackage{tikz}
\usepackage{IEEEtrantools}
\usepackage{anyfontsize}
\usetikzlibrary{shadows.blur}
\usetikzlibrary{arrows.meta}

\definecolor{astFrobenius}{RGB}{216,199,232}
\definecolor{astIntermediate}{RGB}{238,190,126}
\definecolor{astOperator}{RGB}{166,203,178}
\definecolor{guide}{RGB}{125,125,125}

\usepackage[colorlinks=true, urlcolor=violet, linkcolor=blue, citecolor=red, hyperindex=true, linktocpage=true]{hyperref}
\usepackage[capitalise,compress]{cleveref}

\allowdisplaybreaks

\newtheorem{thm}{Theorem}

\makeatletter
\renewcommand{\p@subsection}{}
\renewcommand{\p@subsubsection}{}
\makeatother

\newcommand\bea{\begin{eqnarray}}
\newcommand\eea{\end{eqnarray}}
\newcommand\be{\begin{equation}}
\newcommand\ee{\end{equation}}
\newcommand\bes{\begin{subequations}}
\newcommand\ees{\end{subequations}}
\newcommand\bed{\begin{displaymath}}
\newcommand\eed{\end{displaymath}}
\newcommand\beal{\begin{aligned}}
\newcommand\eeal{\end{aligned}}
\newcommand\bew{\begin{widetext}}
\newcommand\eew{\end{widetext}}
\newcommand\beit{\begin{itemize}}
\newcommand\eeit{\end{itemize}}
\def\bea{\begin{array}}
\def\eea{\end{array}}
\newcommand\been{\begin{enumerate}}
\newcommand\eeen{\end{enumerate}}

\usepackage{verbatim}

\newcommand{\0}{\ket{\bar{0}}}

\newcommand{\an}{{\rm anc}}
\renewcommand{\ap}{{\rm anc'}}

\newcommand{\id}{\mathbbm{1}}

\renewcommand{\H}{\mathcal{H}}
\newtheorem{cor}{Corollary}

\theoremstyle{definition}

\usetikzlibrary{quantikz2}

\definecolor{red1}{rgb}{0.76, 0.23, 0.13}
\definecolor{green1}{rgb}{0.0, 0.5, 0.0}
\usepackage{pifont} 

\usepackage{booktabs}

\newlength{\gap}
\let\abs\relax
\let\norm\relax
\DeclarePairedDelimiterX\abs[1]{\lvert}{\rvert}{#1}%
\DeclarePairedDelimiterX\norm[1]{\lvert}{\rvert}{\hspace{\gap}\delimsize\lvert #1 \delimsize\rvert \hspace{\gap}}
\DeclarePairedDelimiterX\unnorm[1]{\lvert}{\rvert}{\hspace{\gap}\delimsize\lvert \hspace{\gap}\delimsize\lvert #1 \delimsize\rvert \hspace{\gap} \delimsize\rvert \hspace{\gap}}

\makeatletter
\let\oldabs\abs
\def\abs{\@ifstar{\oldabs}{\oldabs*}}
\let\oldnorm\norm
\def\norm{\@ifstar{\oldnorm}{\oldnorm*}}
\let\oldunnorm\unnorm
\def\unnorm{\@ifstar{\oldunnorm}{\oldunnorm*}}
\makeatother

\usepackage[T1]{fontenc}
\usepackage{multirow}
\usepackage{caption}
\usepackage[normalem]{ulem}
\usepackage{subcaption}
\usepackage{soul}

\newcommand{\inlinesection}[1]{\emph{#1}.---}

\begin{document}

\title{Thermal Throttling of Quantum State Transfer}
\author{Twesh Upadhyaya}
\email{tweshu@umd.edu}
\affiliation{Joint Center for Quantum Information and Computer Science, University of Maryland and NIST, College Park, MD 20742, USA}
\affiliation{Department of Physics, University of Maryland, College Park, MD 20742, USA}
\affiliation{Joint Quantum Institute, University of Maryland and NIST, College Park, MD 20742, USA}

\author{T. C. Mooney}
\affiliation{Joint Center for Quantum Information and Computer Science, University of Maryland and NIST, College Park, MD 20742, USA}
\affiliation{Joint Quantum Institute, University of Maryland and NIST, College Park, MD 20742, USA}

\author{Yifan Hong}
\email[Current affiliation: NVIDIA]{}
\affiliation{Joint Center for Quantum Information and Computer Science, University of Maryland and NIST, College Park, MD 20742, USA}
\affiliation{Joint Quantum Institute, University of Maryland and NIST, College Park, MD 20742, USA}

\author{Alexey V. Gorshkov}
\affiliation{Joint Center for Quantum Information and Computer Science, University of Maryland and NIST, College Park, MD 20742, USA}
\affiliation{Joint Quantum Institute, University of Maryland and NIST, College Park, MD 20742, USA}

\begin{abstract}
    Quantum state transfer on qubit lattices is a crucial step in a panoply of quantum information processing tasks. Understanding its fundamental limits in the presence of practical imperfections remains a pressing open question. In particular, it is unclear how thermal noise in the intermediate ancilla sites impedes transfer. In this work, we derive tight lower bounds on the necessary growth of commutator norms to achieve approximate state transfer. Specializing to 1D power-law systems with thermal ancilla states, we demonstrate that state transfer runtimes depend sensitively on the scaling of temperature with system size, improving logarithmic bounds to algebraic ones.
    Our work extends previous results on exact state transfer to the qualitatively different and more practically relevant regime of approximate state transfer. 
\end{abstract}

\maketitle

\begin{figure}[t]
    \centering
\begin{tikzpicture}[scale=0.7]

\def\a{1.6}        
\def\Rblue{0.48}   

\definecolor{softblue}{RGB}{110,155,205}
\definecolor{softred}{RGB}{215,115,115}
\definecolor{softgreen}{RGB}{120,190,145} 

\tikzset{
    bluequbit/.style={
        fill=softblue,
        opacity=0.85
    }
}

\newcommand{\psiQubit}[2]{

    \path[fill=softgreen, opacity=0.85] (#1,#2) circle (\Rblue);
    \foreach \k in {1,...,12} {
        \pgfmathsetmacro{\rk}{\Rblue * \k / 12}
        \pgfmathsetmacro{\ok}{0.75 * exp(-0.25*\k)}
        \path[fill=white, opacity=\ok] (#1,#2) circle (\rk);
    }
    \node at (#1,#2) {$\ket{\psi}$};
}

\psiQubit{0}{0}

\foreach \x/\y in {\a/0, 2*\a/0, 4*\a/0,5*\a/0} {
    \foreach \k in {1,...,12} {
        \pgfmathsetmacro{\rk}{\Rblue * \k / 12}
        \pgfmathsetmacro{\ok}{0.8 * exp(-0.15*\k)}
        \path[fill=softred, opacity=\ok]
            (\x,\y) circle (\rk);
    }
}

\node[opacity=0.55,scale=0.8] at (3*\a,0) {\Huge $\cdots$};

\coordinate (qBL) at (0,0);
\coordinate (qTR) at (5.1*\a,0);

\coordinate (ctrlA) at (2*\a, 1*\a);
\coordinate (ctrlB) at (3*\a, 1*\a);

\tikzset{transferarrow/.style={
    -{Triangle[length=8pt,width=9pt]},
    line cap=round, line join=round
}}

\draw[transferarrow,
      line width=3pt, black!20, opacity=1,
      shorten <=\Rblue cm, shorten >=\Rblue cm]
  ([xshift=2.6pt,yshift=-2.6pt]qBL)
    .. controls ([xshift=2.6pt,yshift=-2.6pt]ctrlA)
    and       ([xshift=2.6pt,yshift=-2.6pt]ctrlB)
    .. ([xshift=2.6pt,yshift=-2.6pt]qTR);

\draw[transferarrow, draw=softgreen!80!black,
      line width=2.4pt, opacity=1,
      shorten <=\Rblue cm, shorten >=\Rblue cm]
  (qBL) .. controls (ctrlA) and (ctrlB) .. (qTR);
\end{tikzpicture}
\caption{Consider transferring a quantum state over lattice sites that are imperfectly initialized, e.g. in thermal states.
}
    \label{fig:keyimage}
\end{figure}
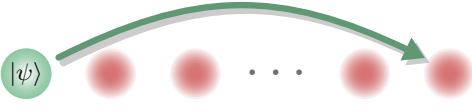

Understanding the speed at which quantum states can be transferred is of fundamental interest in quantum information and helps benchmark subroutines in quantum algorithms, communication, and error correction~\cite{Bose_2003,Christandl2004Perfect,Christandl2005Perfect,Yung2005Perfect,Franco2008Perfect}. 
Specifically, one is often interested in the scenario where, using local interactions, an unknown quantum state on one site of a lattice is to be transported over \emph{ancilla} sites to some final location (Fig.~\ref{fig:keyimage}). In a wide variety of geometries, with interactions ranging from nearest-neighbor \cite{Lieb_1972, hastings2010locality,Wang_2020, Chen_2021} to power-law \cite{Chen_2019_1D, Kuwahara_2020, Tran_2020_hierarchy, Tran2021lrb}, light cones have been derived and can be used to obtain minimum runtimes for state transfer~\cite{Bravyi_2006_LRB,Nachtergaele_2006}. These light cones capture the ultimate speed limits on quantum information spreading and guide the construction of rapid quantum state-transfer protocols.

The state transfer scenario is typically instantiated with ancilla qubits in known quantum states.
But real devices have noisy state preparation~\cite{Tsomokos_2007,Yao2011}, making the effect of ancilla-initialization error on state transfer times important to understand. 
For \emph{exact} state transfer, we recently showed that the robustness of a state transfer protocol is quantified by the dimension of the subspace of pure ancilla states it works on, and that increased robustness necessitates longer minimum runtimes~\cite{Upadhyaya2026robust}. 

In practice, the noise model that is perhaps most ubiquitous is thermalizing noise~\cite{Johnson2022}, whereby each ancilla state is in a mixed thermal state, not a pure state. But for mixed ancilla states, the behavior of exact state transfer is pathological: achieving exact state transfer on \emph{any} full-rank ancilla state necessitates the same minimum time~\cite{Upadhyaya2026robust}. On the other hand, intuitively, we expect that state transfer on a nearly-pure mixed state should be about as hard as on that pure state, not as on the maximally mixed state. 

In this work, we resolve this tension by analyzing the performance of \emph{approximate} state transfer.
Namely, we allow for a nonzero error in the transferred state \cite{Venuti2007,Godsil2012,Sousa2014}, in contrast to many results that are restricted to exactly zero error~\cite{Christandl2004Perfect, Eldredge_2017, Tran2021a, friedman2023,  Chen2023,Hong2024,Upadhyaya2026robust}.
Not only is approximate state transfer a more well-conditioned task, it also 
better captures the open-system dynamics of actual state transfer experiments, where some imprecision is unavoidable. 

Roughly speaking, we prove that the more ``mixed'' the initial ancilla state is, the slower state transfer gets. More precisely, we prove tight lower bounds on the commutator Schatten $p$-norms for any approximate, robust state transfer protocol.
Here, the commutator is taken between an operator on the initial site and a time-evolved operator on the final site.
Our bounds decrease with approximation error and as the spectrum becomes more concentrated onto a single nonzero eigenvalue. The bounds are independent of the specific geometry or interaction; combined with light cones for a specific system, they give minimum runtimes.
The seemingly innocuous change from zero to nonzero error necessitates a new approach to deriving bounds. 
The error on a mixture of states can be strictly lower than that mixture of the individual errors. Thus, the analysis of a set of pure states with specified errors for each and of a mixed state with a single specified error can no longer be treated on the same footing. We prove bounds for both scenarios by generalizing the notion of stabilizer unitaries from \cite{Hong2024} to stabilizer contractions 
and leveraging tools from the theory of convex optimization.
Our bounds recover the results of \cite{Upadhyaya2026robust} at zero error, but reveal a much richer landscape at nonzero error.
We present saturating protocols for our commutator-norm bounds, cementing their tightness.

In 1D power-law systems~\cite{Saffman2010Quantum, Doherty2013Nitrogen, Gadway2016Strongly,Monroe2021Programmable}, where state transfer and light cones have recently attracted much attention~\cite{Hastings_2006, FossFeig_2015, Matsuta_2016, Eldredge_2017, Tran_2019, Chen_2019, Luitz_2019, Else_2020, Chen_2019_1D, Kuwahara_2020,Guo_2020, Tran_2020_hierarchy, Tran2021lrb,  Tran2021a, Hong_2021_fast,Gong_2023}, we apply our lower bounds to derive minimum runtimes for thermal ancilla states. In certain regimes, our varying $p$ bounds are exponentially stronger than those with static $p$---the overwhelming majority of prior work has focused on the special cases of $p=\infty,2$ \cite{hastings2010locality, Chen2023}. 
Letting the temperature scale with system size, we discover a phase transition for when sublinear state transfer is possible. We present an explicit protocol that achieves near-optimal runtimes in certain regimes.

\inlinesection{Approximate state transfer}We consider a lattice of $L$ qubits in arbitrary spatial dimensions with a distance measure between sites. While we focus on qubits, our results directly extend to qudits.
Let $\H_i$ denote the initial Hilbert space, on which an unknown state is prepared. Let $\H_f$ be the final space of the same dimension to which the state is to be transferred. Let $\H_\an=\H_f\otimes \H_a$ be the Hilbert space for the ancillary lattice sites over which the state is transported. We also use the shorthand $\H_\ap=\H_i \otimes \H_a$.

A state transfer protocol can be modeled by a channel $\Xi$ on the composite Hilbert space $\H_i \otimes \H_\an $. This channel could be unitary, and realized by a time-dependent Hamiltonian, or non-unitary, e.g. effected by mid-circuit measurements and feedback or open-system dynamics \cite{friedman2023}. For any non-unitary channel $\Xi$, one can always Stinespring dilate it to a unitary on a larger space and treat the dilating qubits as additional ancillas.
For an ancilla state $\sigma_\an$, the effective transfer channel $\Lambda_{\Xi,\sigma_\an}:L(\H_i) \rightarrow L(\H_f) $ is defined as
\begin{equation}\label{stdefn}
    \Lambda_{\Xi,\sigma_\an}(\rho_i)\coloneqq\Tr_\ap[ \Xi (\rho_i \otimes \sigma_\an) ].
\end{equation}

For an exact protocol, $\Lambda$ is the identity channel. For an approximate protocol, we want to quantify how much $\Lambda$ deviates from the identity channel. A standard error metric for this is the diamond norm, which is well motivated practically and theoretically~\cite{Watrous2018}. It captures the worst case over all input states, including states entangled with a reference system that is untouched by $\Lambda$ but can be used for discrimination. So, $\Xi$ achieves $\epsilon$-approximate state transfer on the ancilla state $\sigma$ if
\begin{equation}\label{approxstdefn}
\tfrac{1}{2}\unnorm{\Lambda_{\Xi,\sigma}-\mathds{1}}_\diamond \leq\epsilon, 
\end{equation}
where $\mathds{1}$ is the identity channel~\footnote{We reserve 2 bars to denote renormalized norms.}. $\epsilon=0$ recovers the usual definition of exact state transfer. We prove some properties of the set of ancilla states where the protocol achieves a given $\epsilon$ in the Supplemental Material (SM)~\cite{suppmat}. From hereon, we generally suppress subscript labels.

\inlinesection{Norms}For a complex vector of length $N$, the $p$-norm ($p\geq 1$) is
\begin{equation}
    \unnorm{\bm v}_p \coloneqq \left(\sum_k \abs{v_k}^p \right)^\frac{1}{p},
\end{equation}
and the renormalized $p$-norm is $\norm{\bm v}_p\coloneqq N^{-1/p}\unnorm{\bm v}_p$; $p\rightarrow \infty$ equals $\max_i \abs{v_i}$ in either case.
The Schatten $p$-norm of an operator is defined as the $p$-norm of its singular values~\cite{bhatia_2013}. For $p>q$, $\unnorm{\hat{O}}_q \geq \unnorm{\hat{O}}_p$ but $\norm{\hat{O}}_q \leq \norm{\hat{O}}_p$.

\inlinesection{Pure ancilla states}
We first establish lower bounds on the commutator norms needed for approximate state transfer with pure ancilla states. 
Generalizing the notion of stabilizer unitaries introduced in \cite{Hong2024} and later used in \cite{Upadhyaya2026robust}, we define stabilizer contractions via the Pauli operators and adjoint state transfer channel as 
\begin{align}\label{stabilizers}
    S_x&\coloneqq X_i \Xi^\dagger (X_f),\nonumber \\ 
    S_y&\coloneqq Y_i \Xi^\dagger (Y_f), \quad \text{and} \\
    S_z&\coloneqq Z_i \Xi^\dagger(Z_f). \nonumber
\end{align}
These definitions reduce to those of the stabilizer unitary formalism when $\Xi$ is a unitary channel.
For exact, unitary state transfer, it was shown that a state is a $+1$ eigenvector of (i.e.~stabilized by) both stabilizer unitaries if and only if the protocol works on that state~\cite{Hong2024}. We generalize this result, proving that for approximate and possibly non-unitary transfer, states on which the protocol has small error are ``approximately'' $+1$ eigenvectors of the contractions. 
We denote the real (i.e.~Hermitian) part of an operator by $\Re \hat{O} \coloneqq \frac{1}{2} (\hat{O}+\hat{O}^\dagger)$.
\begin{thm}\label{pureapproxlbtight}
    Suppose $\Xi$ achieves $\epsilon$-approximate state transfer on the ancilla state $\ket{\Phi}_\an$. Then, for any initial state $\ket{\psi}_i$, the stabilizers satisfy
    \begin{align}\label{firstthmeq}
        \abs{{\bra{\psi}_i \bra{\Phi}_\an  S_\eta \ket{\psi}_i\ket{\Phi}_\an} -1}&\leq2\epsilon,
    \end{align}
    and in turn
    \begin{align}\label{secondthmeq}
        {\bra{\psi}_i \bra{\Phi}_\an \Re S_\eta \ket{\psi}_i\ket{\Phi}_\an}&\geq1-2\epsilon,
    \end{align}
    for $\eta=x,y,z$.
\end{thm}
This theorem validates the intuitive expectation that approximate state transfer leads to approximate stabilization. We prove the theorem and show tightness in the End Matter (EM). 
For $\epsilon\geq1/2$, Eq.~\eqref{secondthmeq} becomes vacuous. 
That the stabilizer conditions are necessary is the direction we need for deriving commutator-norm lower bounds. 
In the SM, we show that the conditions are also sufficient to certify $4\sqrt{3} \epsilon$-approximate state transfer.

We now leverage this theorem to lower bound the commutator norm of an approximate state transfer protocol, given its performance on a set of pure ancilla states. Such a lower bound is useful for the following reason: The norm of $[\Xi^\dagger(X_f), Z_i]$ is upper bounded by light cones for the system under consideration. Hence, a lower bound on the commutator implies a minimum runtime for state transfer. By Eq.~\eqref{stabilizers}, we can alternatively express the commutator as $[X_i S_x, Z_i]$. We will henceforth focus on $S_x$, but analogous results follow immediately for $S_y, S_z$.
\begin{cor}\label{pureapproxcor}
Let $\ket{\Phi_k}$ be a basis of ancilla states with $\Xi$ achieving $\epsilon_k$-approximate state transfer on $\ket{\Phi_k}$. Let $\bm{w}$ be the vector defined by $w_k=\max\{0,1-2\epsilon_k\}$. Then, for any $p\geq1$,
\begin{align}
    \norm{[X_i S_x, Z_i]}_p \geq 2 \norm{\bm w}_{ p}.
\end{align}    
\end{cor}
This corollary is tight. In the SM we provide the proof, a channel $\Xi$ that exactly saturates the bound, and a unitary channel that achieves the bound to within a small constant factor of $2^{1/p}$.
This corollary establishes that allowing for worse approximations makes state transfer easier, i.e. the commutator norm does not have to be grown as large. Note that this result can be applied even if the performance of $\Xi$ is only known on some states. For states where the protocol is not characterized, we can simply set the corresponding $\epsilon_k=1$. 
For example, suppose $\Xi$ achieves $\epsilon\leq1/2$ on each of a set of $S$ orthonormal ancilla states. The corollary gives 
\begin{equation}
     \norm{[X_i S_x, Z_i]}_p \geq 2\cdot 2^{(-L+1)/p} S^{1/p} (1-2\epsilon).
\end{equation}
This gives a quantitative tradeoff between the error and norm that recovers Theorem 1 of \cite{Upadhyaya2026robust} at $\epsilon=0$, with $S$ playing the role of the robust-subspace dimension. 
However, unlike for exact state transfer, this pure-ancilla-state result does not immediately connect to mixed ancilla states, as we now expound upon.

\inlinesection{Mixed ancilla states}A natural and ubiquitous model for noise in the ancilla state is thermal noise~\cite{Johnson2022}. This is especially relevant in solid-state architectures, which typically have higher noise levels~\cite{Schulman1998,Yao2011,Tsomokos_2007}. 
Exact state transfer on mixed ancilla states is poorly conditioned in the following sense: In a 1D power-law system with the initial ancilla state $(1-\Delta) \dyad{00...0}+\frac{\Delta}{2^{L-1}} \id$, for $\Delta>0$, any protocol that achieves exact state transfer on this state must be state-independent and therefore take linear time for $\alpha>2$~\cite{Upadhyaya2026robust,Tran_2020_hierarchy,Chen2021}. But, when $\Delta=0$, exact state transfer can be achieved via a fast Greenberger-Horne-Zeilinger (GHZ) protocol in sublinear time for $\alpha<3$~\cite{Tran2021a}. Therefore, the minimum runtime of exact state transfer is not continuous with respect to changes in the initial ancilla state. This motivates us to consider approximate state transfer as a more well-conditioned physical task.
For $\Delta>0$, the fast GHZ protocol achieves, in sublinear time, $\Delta$-approximate state transfer by Lemma~S3 in the SM. Hence, there can be a large separation between the runtimes for approximate and exact state transfer. 

Recall that, for exact state transfer, a protocol works on a mixed state if and only if it works on every pure state in the mixture~\cite{Upadhyaya2026robust}.
This reduction does not extend to approximate transfer. Namely, a protocol can achieve parametrically lower error on a mixture of two pure states than on either of them alone. For example, consider perfect SWAP followed by small rotation errors $U_\pm\ket{0}=\cos \theta \ket{0} \pm \sin \theta \ket{1}$. The state transfer error is $\epsilon_\text{unitary}\approx \theta$. The error for the channel $\Lambda(\cdot)=\frac{1}{2} U_+ \cdot U_+^\dagger + \frac{1}{2} U_- \cdot U_-^\dagger $ is only $\epsilon_\text{mixed}\approx\theta^2$.
(This is reminiscent of results on decreasing Trotter error in quantum simulations by using mixed-unitary dynamics~\cite{Childs2019fasterquantum}.)
Despite this effect, in this section we bound commutator norms given only the promise that a protocol achieves $\epsilon$-approximate state transfer on a particular mixed state $\sigma=\sum_{k=1}^r \lambda_k \dyad{\Phi_k}$. 

Our main theorem for approximate state transfer on mixed ancilla states is as follows:
\begin{thm}\label{thmmixed}
    Suppose $\Xi$ achieves $\epsilon$-approximate state transfer ($\epsilon\leq\frac{1}{2}$) on a mixed ancilla state $\sigma$ with spectrum $\bm \lambda$. Then, the commutator norm must satisfy
\begin{equation}
 \norm{[X_i S_x, Z_i]}_p \geq 2\norm{\bm y}_p,
\end{equation}
where $\bm y$ is as follows for $1<p<\infty$:
\begin{equation}\label{yvec}
    y_k = \min \{1, \lambda' \lambda_k^{\frac{1}{p-1}}\}, 
\end{equation}
with $\lambda'$ determined by
\begin{equation}
    \sum_k \lambda_k \min \{1, {\lambda'} \lambda_k ^{\frac{1}{p-1}}\} = 1-2\epsilon. \label{lambdaconstr}
\end{equation}
\end{thm}
At the edge cases, $\norm{\bm y}_\infty=1-2\epsilon$ and $\norm{\bm y}_1 = 2^{-L+1} [c-1 +(1-2\epsilon-\sum_{k=1}^{c-1} \lambda_k)/ \lambda_c ]$. Here, $c$ is the cutoff where, with the eigenvalues arranged in decreasing order, $\sum_{k=1}^{c-1} \lambda_k<1-2\epsilon$ and $\sum_{k=1}^{c} \lambda_k\geq 1-2\epsilon$ ($\norm{\bm y}_p=0$ when $\epsilon=1/2$).

\begin{cor}\label{cormixed}
    A weaker but closed-form bound is     
    \begin{equation}
    \norm{[X_i S_x, Z_i]}_p \geq 2\cdot 2^{(-L+1)/p} \frac{1}{\unnorm{\bm \lambda }_{\frac{p}{p-1}}} (1 - 2\epsilon),
    \end{equation}    
    with ${\unnorm{\bm \lambda}_{\frac{p}{p-1}}}={\unnorm{\bm \lambda}_\infty}$ at $p=1$.
\end{cor}

As compared to the exact case, Theorem~\ref{thmmixed} has a small $\epsilon$ correction and the rank is replaced by an $\epsilon$-tolerant approximation to it. We prove Theorem~\ref{thmmixed} is tight by providing a saturating state-transfer protocol in the SM. In analogy with Theorem 1 of \cite{Upadhyaya2026robust}, $\unnorm{\bm y}_p^p$ plays the role of an \emph{effective subspace dimension} (though it may be non-integral).
At fixed $\epsilon$ and $p$, larger $\lambda'$ arise from less concentrated spectra, as we discuss in more detail in the SM.
The theorem and corollary bounds are both increasing with $p$, as we prove in the SM. Combined with the fact that light cones become weaker at larger $p$ \cite{Chen2021}, we obtain a tradeoff that leads to a nontrivial optimization over $p$, similar to the result for robust exact state transfer~\cite{Upadhyaya2026robust}.

We now examine several specific cases. First, suppose there are $S$ nonzero eigenvalues and $\epsilon=0$. Then, to satisfy Eq.~\eqref{lambdaconstr}, we must have $y_k=1$ for every nonzero $\lambda_k$, so we get $\norm{[X_i S_x, Z_i]}_p \geq 2\cdot 2^{(-L+1)/p} S^{1/p}$, recovering the result of~\cite{Upadhyaya2026robust}.
Suppose the $S$ eigenvalues are uniform and $\epsilon$ is arbitrary. By symmetry, we must have that $\lambda' = {(1 - 2\epsilon)}/{\sum_k \lambda_k^{\frac{p}{p-1}}}$ so that 
\begin{align}
\unnorm{\bm y}_p&=\frac{1 - 2\epsilon}{\sum_k \lambda_k^{\frac{p}{p-1}}}  \left( \sum_k \lambda_k^{\frac{p}{p-1}} \right)^{1/p}\\
    &=(1 - 2\epsilon)\left( \sum_k \lambda_k^{\frac{p}{p-1}} \right)^\frac{1-p}{p}\\
    &=\frac{1 - 2\epsilon}{\unnorm{\bm \lambda}_{\frac{p}{p-1}}},
\end{align}
matching the bound from the corollary. 
Next, when $p\rightarrow\infty$, $\lambda'\rightarrow 1-2\epsilon$ to satisfy the constraint. Then, $2\norm{\bm y}_p\rightarrow 2 \norm{\bm y}_\infty=2(1-2\epsilon)$. This matches the corollary's bound, as $\frac{p}{p-1}\rightarrow 1$. 
At $p=2$, the corollary indicates that states with lower purity necessitate higher commutator norms, in line with the intuition from our opening example. For $\epsilon\geq\frac{1}{2}$, we have $\lambda'=0$ and thus a trivial bound. For $\epsilon=0$, $\lambda'={\lambda_\text{min}^{1/(1-p)}}$ and the theorem bound only counts the nonzero eigenvalues, while the bound in the corollary depends on the specific values of those nonzero eigenvalues.
Thus, we expect the corollary and theorem to be close in three regimes: when the spectrum $\bm \lambda$ is close to uniform, when $p$ is large, and when $\epsilon$ is not very small. In the SM we validate this expectation numerically.

\inlinesection{1D power-law systems}We can combine Theorem~\ref{thmmixed} or Corollary~\ref{cormixed} with power-law light cones to derive bounds on state transfer in such systems. Specifically, we focus on the transfer of a single qubit via unitaries generated by power-law Hamiltonians.
For $p=\infty,2$, tight light cones are known~\cite{Tran_2020_hierarchy,Chen2021,Kuwahara2021,Kuwahara_2020,Tran2021lrb}. Though not tight, the following light cone is known for all $p\geq 2$ \cite{Chen2021}. With $\alpha$ denoting the power-law exponent and for sites separated by a distance $L$,
\begin{equation}\label{pnormlightcone}
    \norm{[U^\dagger X_f U,Z_i]  }_p\geq c  \implies t \gtrsim c  \frac{R(L)}{ \sqrt{p}}, 
\end{equation}
with
\begin{equation}\label{eq:R(r)}
    R(L)=\begin{cases}
    L^{\alpha-3/2} &\quad 3/2<\alpha < 5/2\\
    L/\ln^{3/2}(L) &\quad \alpha = 5/2\\
    L &\quad \alpha>5/2.
    \end{cases}
\end{equation}

So, in a 1D power-law system with $\alpha>3/2$, a runtime bound for
$\epsilon$-approximate state transfer ($\epsilon\leq\frac{1}{2}$) over a mixed ancilla state $\sigma_\an$ with spectrum $\bm \lambda$ can be derived by plugging Eq.~\eqref{pnormlightcone} into Corollary~\ref{cormixed},
\begin{equation}\label{unoptimizedbd}
    t\gtrsim 2(1-2\epsilon){R(L)} \frac{1}{\sqrt{p}} 2^{(-L+1)/p} \frac{1}{\unnorm{\bm \lambda }_{\frac{p}{p-1}}}.
\end{equation}
The best such bound is then found by optimizing over $p$. We solve this optimization in the following theorem.

\begin{thm}\label{thmpopt}
For a probability distribution $\bm r$, let $\bm r(q)$ be the probability distribution defined by exponentiating elementwise, $r_k \rightarrow r_k^q$, then renormalizing. Let $H(\bm r)$ be the Shannon entropy. Then, 
Eq.~\eqref{unoptimizedbd} is maximized at the unique $p^*$ satisfying
\begin{equation}\label{eqthmpopt}
\frac{p^*}{2} + H\left[\bm \lambda\left(\frac{p^*}{p^*-1}\right)\right]=(L-1) \ln 2, 
\end{equation}
when $1 + H\left[\bm\lambda\left(2\right)\right]< (L-1) \ln 2$, and $p^*=2$ otherwise.
\end{thm}

We discuss some implications of Theorem~\ref{thmpopt}; the proof can be found in the SM. At least one of $p$ or $H$ has to be extensive in $L$. 
The entropy term in Eq.~\eqref{eqthmpopt} increases with $p$ and at $p=\infty$ is simply the entropy of $\bm \lambda$ (see the SM). So if $H(\bm \lambda)=o(L)$, for example for a pure ancilla state, then for large $L$ we must have $p^*\approx2 \ln 2 (L-1) $. 
If $\bm \lambda$ is the uniform distribution over $S$ values, then $\bm \lambda(\frac{p}{p-1})=\bm \lambda$ and the entropy term is simply $\ln S$. Then at large enough $L$, $p^*\approx2((L-1) \ln 2 - \ln S)$. 
This theorem proves that for a more ``mixed'' ancilla state,
the optimal $p$ for state transfer runtime bounds is smaller.

We now have the machinery in place to study approximate state transfer with each ancilla qubit initialized in the same thermal state, 
\begin{equation}\label{thermalbackground}
\sigma_\an=\begin{pmatrix} \frac{1}{2}(1+\delta) & 0 \\0 & \frac{1}{2}(1-\delta) \end{pmatrix}^{\otimes L-1} \, ,    
\end{equation}
where we allow the gap $\delta>0$ to depend on $L$. 
We have used Corollary~\ref{cormixed} rather than Theorem~\ref{thmmixed} as the input to Theorem~\ref{thmpopt} to enable analytic optimization over $p$. For small $\delta$, we do not expect this to materially change the runtime bounds (see discussion above).

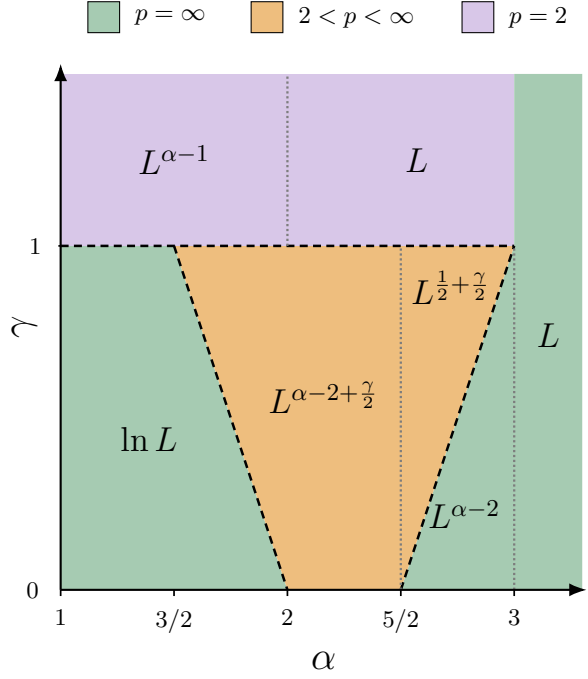
\begin{figure}
    \centering
   \begin{tikzpicture}[x=3.00cm,y=4.55cm,>=Latex,scalinglabel/.style={font=\fontsize{13}{15}\selectfont},axislabel/.style={font=\fontsize{15}{17}\selectfont}]

\fill[astOperator] (1.0,0) -- (2.0,0) -- (1.5,1.0) -- (1.0,1.0) -- cycle;
\fill[astOperator] (2.5,0) -- (3.0,1.0) -- (3.0,1.50) -- (3.3,1.50) -- (3.3,0) -- cycle;

\fill[astIntermediate]
  (1.5,1.0) -- (3.0,1.0) -- (2.5,0) -- (2.0,0) -- cycle;

\fill[astFrobenius] (1.0,1.0) rectangle (3.0,1.50);

\draw[densely dashed,line width=0.95pt] (1.5,1.0) -- (2.0,0.0);
\draw[densely dashed,line width=0.95pt] (2.5,0.0) -- (3.0,1.0);
\draw[densely dashed,line width=0.95pt] (1.0,1.0) -- (3.0,1.0);

\draw[guide,densely dotted,line width=0.85pt] (2.0,1.0) -- (2.0,1.50);
\draw[guide,densely dotted,line width=0.85pt] (2.5,0) -- (2.5,1.0);
\draw[guide,densely dotted,line width=0.85pt] (3.0,0) -- (3.0,1.0);

\draw[line width=0.8pt,->] (1.0,0) -- (3.32,0);
\draw[line width=0.8pt,->] (1.0,0) -- (1.0,1.53);

\foreach \x/\lab in {1/{1},1.5/{3/2},2/{2},2.5/{5/2},3/{3}} {
  \draw[line width=0.65pt] (\x,0) -- ++(0,-0.018);
  \node[below=4pt] at (\x,0) {$\lab$};
}
\node[left=5pt] at (1.0,0) {$0$};
\draw[line width=0.65pt] (1.0,1) -- ++(-0.010,0);
\node[left=4pt] at (1.0,1) {$1$};

\node[axislabel,below=20pt] at (2.16,0) {$\alpha$};
\node[axislabel,rotate=90] at (0.84,0.765) {$\gamma$};

\node[scalinglabel] at (1.50,1.25)
  {$L^{\alpha-1}$};

\node[scalinglabel] at (2.16,0.56)
  {$L^{\alpha-2+\frac{\gamma}{2}}$};
\node[scalinglabel,inner sep=0pt] at (2.72,0.88)
  {$L^{\frac{1}{2}+\frac{\gamma}{2}}$};
\node[scalinglabel] at (1.39,0.44)
  {$\ln L$};
\node[scalinglabel] at (2.78,0.22)
  {$L^{\alpha-2}$};

\node[scalinglabel] at (3.15,0.74)
  {$L$};
\node[scalinglabel] at (2.57,1.25)
  {$L$};

\begin{scope}[shift={(1.18,1.62)}]
    \fill[astOperator] (0-0.06,0) rectangle (0.14-0.06,0.09);
    \draw (0-0.06,0) rectangle (0.14-0.06,0.09);
    \node[anchor=west] at (0.17-0.06,0.045) {$p=\infty$};

    \fill[astIntermediate] (0.72-0.06,0) rectangle (0.86-0.06,0.09);
    \draw (0.72-0.06,0) rectangle (0.86-0.06,0.09);
    \node[anchor=west] at (0.89-0.06,0.045) {$2<p<\infty$};

    \fill[astFrobenius] (1.76-0.17,0) rectangle (1.90-0.17,0.09);
    \draw (1.76-0.17,0) rectangle (1.90-0.17,0.09);
    \node[anchor=west] at (1.93-0.17,0.045) {$p=2$};
\end{scope}
\end{tikzpicture}
    \caption{``Phase diagram'' of the scaling of runtime lower bounds for approximate state transfer over thermal ancilla states in 1D power-law systems. Power-law exponent is $\alpha$, $\epsilon$ is constant, and the thermal ancilla state is specified by $\delta=L^{-\gamma/2}$. (If $\epsilon$ depends on $L$, a factor of $1-2\epsilon(L)$ scales the bounds in all regions.) Each region is labeled by the strongest runtime bound (omitting all subleading polylog factors), and its color (see legend at top of figure) indicates which light cone is used in combination with our commutator bounds. The operator-norm ($p=\infty$) bound is nontrivial over the entire region, but is superseded in certain regimes by the Frobenius ($p=2$) or general $p$-norm bounds.
    Prior to our work, only the Frobenius bounds in the infinite-temperature ($\gamma\rightarrow\infty$) limit and the operator-norm bounds were known. }
    \label{phasediagram}
\end{figure}

\begin{cor}\label{thermalcor}
Consider a 1D power-law system with $\alpha>3/2$. Let each ancilla qubit be in a thermal ancilla state with gap $\delta(L)$ (Eq.~\eqref{thermalbackground}) such that $\delta(L)=o(1)$ and $\delta(L)=\omega(1/\sqrt{L})$. Then, for large $L$, any $\epsilon$-approximate ($\epsilon\leq\frac{1}{2}$) unitary state transfer protocol  (Eq.~\eqref{approxstdefn}) on this ancilla state requires time
\begin{align}
    t \gtrsim     (1 - 2\epsilon) \frac{R(L)}{\delta \sqrt{L}}.
\end{align}
\end{cor}
This corollary establishes that higher-temperature intermediate states slow down state transfer.
We provide the proof and bounds for other regimes of $\alpha,\delta$ in the EM.
The runtime lower bounds, coming from our commutator bounds combined with the best light cones, as derived in Corollary~\ref{thermalcor} and the EM, are summarized in Figure~\ref{phasediagram}. At higher temperatures, smaller $p$ give better bounds on approximate state transfer.

Our results inform the design of future quantum platforms. 
As devices are scaled up, each qubit may be more difficult to initialize.
For fast state transfer subroutines to be achievable, this is fine as long as the temperature of each qubit does not increase too quickly with the system size.
For example, for $2<\alpha<5/2$, our bounds rule out sublinear state transfer unless $\delta\gtrsim\frac{1}{\sqrt{L}}$.

We now illustrate the importance of optimizing over $p$ in our bounds.
Say $3/2<\alpha<5/2$ and $\delta=\frac{1}{L^{\gamma/2}}$ with 
$1>\gamma>4-2\alpha$. Our bound from Corollary~\ref{thermalcor} is $t\gtrsim L^{\alpha-2+\gamma/2}$. 
On the other hand, the bound from Theorem~\ref{thmmixed} and 
the operator-norm light cone is $t\gtrsim \ln L $ ($3/2<\alpha<2$) or $t\gtrsim L^{\alpha-2}$ ($2<\alpha<5/2$) which is weaker than the optimized-$p$ bound. The bound from Theorem~\ref{thmmixed} and the Frobenius-norm light cone tends to zero at large $L$. (See the SM for the proofs.)
Thus, our optimal-$p$ bounds can be parametrically stronger than static-$p$ ones. \emph{This is even though the ancilla state is almost maximally mixed.}

\inlinesection{Cooling protocol}We now present a protocol for achieving approximate state transfer over thermal ancilla states using power-law dynamics. Its runtime approaches our lower bound in certain regimes. The protocol proceeds in two steps: We use schemes based on classical error correction to probabilistically reset a spaced sublattice of qubits to $\ket{0}$; then, on this sublattice, we use the fast GHZ protocol of \cite{Tran2021a}.

We desire a block size $B$ and a reset unitary acting on each block satisfying two properties: The reset step should be faster than the fast GHZ step and the probability of the reset succeeding on \emph{all} blocks should be constant.
There are different strategies to perform the reset unitarily and in place. The first is to use a circuit for reversible majority vote (assuming $B$ is odd) and the second is to use algorithmic cooling~\cite{Schulman1998}.

For majority vote to succeed on all blocks with nondecaying probability, it suffices that $\frac{L}{B}\exp{-\frac{1}{2} B \delta^2}\rightarrow 0$ by the Chernoff bound. With $\delta=\frac{1}{L^{\gamma/2}}$ as before, any block size scaling faster than $L^\gamma \ln L$ sends the error to zero. The scaling $B\sim \delta^{-2}$ can also be deduced from Theorem 1 of \cite{Schulman1998} for algorithmic cooling. Majority vote can be realized as one of the outputs of a reversible, symmetric Boolean function when $B$ is odd. To implement it, we can use standard circuit constructions involving SWAPs and pairwise comparisons. These have nearest-neighbor gate depth quadratic in $B$, so can be realized in time $t\sim B^2$ using power-law interactions. This quadratic scaling matches \cite{Schulman1998}. 
The total protocol time is then
\begin{align}
    t = t_\text{reset} + t_\text{rescaled transfer}\\
    \simeq B^2 + B^\alpha t_{GHZ}\left(\frac{L}{B}\right),
\end{align}
where $t_{GHZ}(M)$ is the time to implement the fast GHZ protocol on a lattice of $M$ qubits~\cite{Tran2021a}.
Say $2<\alpha<5/2$; the second term dominates and gives
\begin{equation}
     t \simeq L^{\alpha-2+2\gamma},
\end{equation}
up to polylog factors, compared to our above lower bound
\begin{equation}
     t \simeq  L^{\alpha-2+\gamma/2}.
\end{equation}
The two differ by $L^{3\gamma/2}$; finding protocols to close this gap is an interesting future direction.

When $\delta$ is a nonzero constant, a block size of $B=\ln^2 L$ suffices and leads to a protocol runtime that is polylogarithmic for $1<\alpha<2$. The bound from Theorem~\ref{thmmixed} and the operator-norm light cone is also polylogarithmic.

\inlinesection{Outlook}Our work provides a comprehensive picture of state transfer over noisy ancilla states. For a target error over a mixed state, we prove tight lower bounds on the growth of Heisenberg-evolved commutator norms. In 1D power-law systems, our results establish that slow-scaling thermal gaps bottleneck information propagation.

That our bounds are saturated indicates they cannot be improved further. Yet, they only depend on the ancilla state's spectrum. This suggests commutator norms are fundamentally limited for constraining state transfer. 
Understanding how state transfer performance is affected by commutators' eigenvectors, for example their entanglement structure~\cite{Kessler_2014}, is an exciting avenue to explore.

Improvements are likely also possible at the level of the protocols. Existing protocols in power-law systems largely focus on exact transfer with perfectly initialized ancillas. As we touched upon in this work, speedups are achievable by designing algorithms to tolerate ancilla noise at the cost of a small approximation error. Recent developments for all-to-all interactions may provide inspiration~\cite{Yin_2025_all}.

\inlinesection{Acknowledgments}We thank Chao Yin and Nicole Yunger Halpern for helpful discussions. We were supported in part by the DoE ASCR Quantum Testbed Pathfinder program (award No.~DE-SC0024220), NSF STAQ program, NSF QLCI (award No.~OMA-2120757), ONR MURI, AFOSR MURI, DARPA Agreement HR00112490357, ARL (W911NF-24-2-0107) NQVL:QSTD:Pilot:DLPQC, and NQVL:QSTD:Design:FTL. We also acknowledge support from the U.S.~Department of Energy, Office of Science, Accelerated Research in Quantum Computing, Fundamental Algorithmic Research toward Quantum Utility (FAR-Qu) and from the U.S.~Department of Energy, Office of Science, National Quantum Information Science Research Centers, Quantum Systems Accelerator (award No. DE-SCL0000121).
\vspace{-0.15cm}
\bibliography{mainbib}

\clearpage
\appendix

\onecolumngrid
\center{\normalsize\textbf{End Matter}}
\vspace{1em}
\twocolumngrid
\justifying

\inlinesection{Proof of Theorem~\ref{pureapproxlbtight}}We prove the result for $S_x$; an analogous argument holds for $S_y,S_z$. The key idea is to express stabilizer expectations in terms of the adjoint state transfer map. We first derive the form of the adjoint mapping. Since adjoints preserve inner products, 
\begin{align}
    \hspace{-1.3cm}
    \Tr(A_{f} \Lambda_{\dyad{\Phi}}(B_{i}))&= \Tr_f(A_{f} \Tr_\ap[\Xi (B_{i} \otimes \dyad{\Phi}_\an) ])\\
&=\Tr[(A_{f}\otimes\id_\ap) \Xi (B_{i} \otimes \dyad{\Phi}_\an)  ]\\
    &=\Tr[\Xi^\dagger (A_{f}\otimes\id_\ap) (B_{i} \otimes \dyad{\Phi}_\an)]\\
    &=\Tr_i[\bra{\Phi}_\an \Xi^\dagger (A_{f}\otimes\id_\ap)  \ket{\Phi}_\an  B_{i}],\label{adjointdefn}
\end{align}
so $\Lambda^\dagger_{\dyad{\Phi}}(A_f)= \bra{\Phi}_\an \Xi^\dagger (A_{f}\otimes\id_\ap) \ket{\Phi}_\an$.
From this, we conclude that 
\begin{align}
\bra{\psi} \bra{\Phi} S_x \ket{\psi} \ket{\Phi}&= \bra{\psi} \left( X_i \bra{\Phi}  \Xi^\dagger (X_f)  \ket{\Phi} \right) \ket{\psi}\\
&=\bra{\psi} X_i \Lambda^\dagger_{\dyad{\Phi}}(X_f)\ket{\psi}.
\end{align}

Induced norms for channels are related to those of their adjoints via
\begin{equation}
    \unnorm{\Lambda}_{p\rightarrow q} = \unnorm{\Lambda^\dagger}_{q^*\rightarrow p^*},\label{channelnormdual}
\end{equation}
where $\cdot^*$ indicates the H\"{o}lder dual. Since the diamond norm upper bounds the induced trace norm, $\unnorm{\Lambda_{\dyad{\Phi}}-\mathds{1}}_{1\rightarrow 1} \leq 2\epsilon$ by assumption. \newpage
Thus, $\unnorm{\Lambda^\dagger_{\dyad{\Phi}}-\mathds{1}}_{\infty\rightarrow \infty}\leq 2\epsilon$ which implies
\begin{align}
2\epsilon&\geq \unnorm{\Lambda^\dagger_{\dyad{\Phi}}(X_f)-\mathds{1}(X_f)}_{\infty}\\
&=\unnorm{X_i \Lambda^\dagger_{\dyad{\Phi}}(X_f)-\id}_{\infty}\\
&\geq \abs {\bra{\psi} X_i \Lambda_{\dyad{\Phi}}^\dagger(X_f)\ket{\psi} - 1},
\end{align}
where the first line follows because $\unnorm{X_f}_\infty=1$, the second because the operator norm is unitarily invariant, and the third because the operator norm bounds expectation values. This gives the first part of the theorem.

By Eq.~\eqref{channelnormdual}, as the adjoint of a trace-preserving map, $\unnorm{\Lambda_{\dyad{\Phi}}^\dagger}_{\infty \rightarrow \infty}=1$ so $\Lambda_{\dyad{\Phi}}^\dagger(X_f)$ and thus $S_x$ are contractions. Let $\bra{\psi} X_i \Lambda_{\dyad{\Phi}}^\dagger(X_f)\ket{\psi}=a+bi$. Since this complex number is the expectation value of a contraction, $\abs{a+bi}\leq 1$ so in particular $a\leq1$. We have 
\begin{align}
    2\epsilon &\geq \sqrt{(a-1)^2+b^2}\\
    &\geq 1-a.
\end{align}
Noting that ${\bra{\psi}_i \bra{\Phi}_\an \Re \{ S_x\} \ket{\psi}_i\ket{\Phi}_\an}=\Re {\bra{\psi}_i \bra{\Phi}_\an S_x \ket{\psi}_i\ket{\Phi}_\an}$ completes the proof.

\inlinesection{Saturating Protocol for Theorem~\ref{pureapproxlbtight}}Extend to a basis of ancilla states $\ket{\Phi_k}$ with $\ket{\Phi_1}=\ket{\Phi}$. Define effective Pauli operators on the qubit subspace $\ket{\Phi_1},\ket{\Phi_2}$, e.g. $\tilde{Y}=-i \dyad{\Phi_1}{\Phi_2} +i\dyad{\Phi_2}{\Phi_1}+\sum_{k>2} \dyad{\Phi_k}$.
Consider the unitary 
\begin{equation}
U=\mathrm{SWAP}_{i,f} \left( \sqrt{1-\epsilon} \id_i \otimes \tilde{\id} + i\sqrt{\epsilon} Y_i \otimes \tilde{Y}  \right).  
\end{equation}
The effective channel it realizes is
\begin{widetext}
\begin{align}
   \hspace{-1cm} \Lambda_{U,\dyad{\Phi}}(\rho) &= \Tr_\ap [ \mathrm{SWAP}_{i,f} \left( \sqrt{1-\epsilon} \id_i \otimes \tilde{\id} + i\sqrt{\epsilon} Y_i \otimes \tilde{Y}  \right) \rho_i \otimes \dyad{\Phi}_\an  \left( \sqrt{1-\epsilon} \id_i \otimes \tilde{\id} -i \sqrt{\epsilon} Y_i \otimes \tilde{Y}  \right) \mathrm{SWAP}_{i,f}  ]\\
    &=\Tr_\ap [ (1-\epsilon) \rho_f \otimes \dyad{\Phi}_\ap + {\epsilon} Y\rho_f Y \otimes \tilde{Y}\dyad{\Phi}_\ap \tilde{Y}+ \ldots ]\\
    &= (1-\epsilon) \rho + {\epsilon}Y\rho Y,
\end{align}
\end{widetext}
where in the second line we omit cross terms that are zeroed out by the partial trace. We now verify $U$ achieves $\epsilon$-approximate state transfer. The diamond norm can be computed by acting the transfer channel on part of a possibly larger entangled state $\ket{\psi}_{i i'}$,
\begin{equation}
    \unnorm{\Lambda_{U,\sigma}-\mathds{1}}_\diamond = \max_{\ket{\psi}} \epsilon \unnorm{\dyad{\psi}-Y \dyad{\psi}Y }_1.
\end{equation}
By inspection, $\ket{\psi}=\ket{00}$ achieves the maximum, saturating the upper bound of $\unnorm{\Lambda_{U,\sigma}-\mathds{1}}_\diamond \leq 2\epsilon$ from the triangle inequality.

Next, we compute the stabilizers.
\begin{align}
    \hspace{-4cm} S_x &= X_i \left( \sqrt{1-\epsilon} \id_i \otimes \tilde{\id} -i \sqrt{\epsilon} Y_i \otimes \tilde{Y}\right) (X_i \otimes \tilde{\id}) \\ &\quad \quad\quad\quad  \left( \sqrt{1-\epsilon} \id_i \otimes \tilde{\id} + i\sqrt{\epsilon} Y_i \otimes \tilde{Y}\right)\\
    &= X_i \bigg( (1-\epsilon) X_i \otimes \tilde{\id} -\sqrt{\epsilon}\sqrt{1-\epsilon} Z_i \otimes \tilde{Y} \\
    &\quad \quad -\sqrt{\epsilon}\sqrt{1-\epsilon} Z_i \otimes \tilde{Y} - \epsilon X_i\otimes \tilde{\id}  \bigg)\\
    &= (1-2\epsilon) \id_i \otimes \tilde{\id} + 2i\sqrt{\epsilon}\sqrt{1-\epsilon} Y_i \otimes \tilde{Y}.
\end{align}
Therefore, $\Re S_x=(1-2\epsilon) \id$ so ${\bra{\psi}_i \bra{\Phi}_\an \Re S_x \ket{\psi}_i\ket{\Phi}_\an}=1-2\epsilon$ follows. Note also that $\expval{\tilde{Y}}{\Phi}=0$ so $\abs{\bra{\psi}_i \bra{\Phi}_\an S_x \ket{\psi}_i\ket{\Phi}_\an}=1-2\epsilon$.
Thus the theorem is saturated. Analogous constructions saturate the bounds with $S_y, S_z$.

\inlinesection{Proof of Corollary \ref{thermalcor}}We introduce some notation and identities for the proof of the corollary. The mixed state is a tensor product so $\bm \lambda = (r,1-r)^{\otimes L -1}$. For the thermal state under consideration, $r=\frac{1}{2}(1+\delta)$.
In addition to the binary entropy $h(r)=-r\ln r - (1-r) \ln (1-r)$, we denote the binary entropy of the exponentiated and renormalized distribution as $h(r,q) \coloneqq h( r^q/[r^q+(1-r)^q] ) $. By the additivity of Shannon entropy we have $H(\bm \lambda)= (L-1) h(r)$. Similarly, $\unnorm{\bm \lambda}_p=\unnorm{(r,1-r)}_p^{L-1}$ which is essentially the additivity of R\'{e}nyi entropies. Introducing the normalizing factor on both sides of the previous equation gives $\norm{\bm \lambda}_p=\norm{(r,1-r)}_p^{L-1}$. Elementwise exponentiation behaves nicely with the tensor product---namely, 
\begin{equation}
\bm \lambda(q) = \left[\frac{1}{r^q+(1-r)^q}\begin{pmatrix}
r^q \\
(1-r)^q
\end{pmatrix}\right] ^{\otimes L-1}. 
\end{equation}
This implies 
\begin{equation}\label{qentropyadd}
H(\bm \lambda(q))= (L-1) h(r,q),
\end{equation}
which will be useful later on.

We now proceed to the corollary's proof. From Theorem~\ref{thmpopt}, the critical point condition is
\begin{align}
\frac{p^*}{2} + H\left[\bm \lambda\left(\frac{p^*}{p^*-1}\right)\right]=(L-1) \ln 2.
\end{align}
Using Eq.~\eqref{qentropyadd} and rearranging, the condition is equivalently 
\begin{align}\label{iidoptp}
\frac{p^*}{2(L-1)}  = \ln 2 -   h\left(r,\frac{p^*}{p^*-1}\right).
\end{align}
We are assuming $2r=1+\delta>1$, i.e. the thermal state is at finite temperature, so $h\left(r,\frac{p^*}{p^*-1}\right)\leq h(r) < \ln 2$ (for the first inequality, refer to the proof of Theorem~\ref{thmpopt}). Suppose as $L\rightarrow \infty$, $\delta$ approaches a nonzero constant. Then, for Eq.~\eqref{iidoptp} to be satisfied, $p^* \propto L$. But in the limit of large $L$, $\frac{p^*}{p^*-1}\approx 1$ so 
\begin{equation}\label{entropydeficit}
    p^* \approx 2(L-1)(\ln 2 - h(r)).
\end{equation}
In this case, the best bound is obtained by directly using the operator norm bounds for 1D power-law systems, rather than the general bound evaluated at this $p^*$.

Next consider the case where the gap closes as $L\rightarrow \infty$. Then, we can approximate the binary entropy at small $\delta$:
\begin{align}
\hspace{-0.7cm}    h(r) &= -\frac{1}{2} (1+\delta) \ln(1+\delta) -\frac{1}{2} (1-\delta) \ln(1-\delta) + \ln 2 \\
    &\approx -\frac{1}{2} (1+\delta) \left(\delta-\frac{\delta^2}{2}\right) -\frac{1}{2} (1-\delta) \left(-\delta-\frac{\delta^2}{2}\right) + \ln 2 \\
    &=\ln 2 -\frac{\delta^2}{2}. \label{approxentr}
\end{align}
We can also approximately compute
\begin{align}
\frac{1}{r^2+(1-r)^2}\begin{pmatrix} r^2 \\ (1-r)^2 \end{pmatrix}
&\approx\frac{1}{\frac{1}{4}(1+2\delta)+\frac{1}{4}(1-2\delta)}\begin{pmatrix} \frac{1}{4}(1+2\delta) \\ \frac{1}{4}(1-2\delta) \end{pmatrix}\\
&=\begin{pmatrix} \frac{1}{2}(1+2\delta) \\ \frac{1}{2}(1-2\delta) \end{pmatrix}.
\end{align}
Then, by Eq.~\eqref{approxentr}, $h(r,2)\approx \ln 2 -2\delta^2$. From Eq.~\eqref{iidoptp}, this implies 
\begin{equation}\label{pconstraint}
    \frac{\delta^2}{2} \lesssim \frac{p^*}{2(L-1)}  \lesssim 2\delta^2.
\end{equation}
If $\delta=o({1}/{\sqrt{L}})$, then $p^*$ is driven to zero at large $L$. This is outside the allowed range, so $p^*=2$ as per Theorem~\ref{thmpopt}. 
Then, the specific Frobenius light cone~\cite{Tran_2020_hierarchy,Chen2021,Kuwahara2021} should be used rather than the weaker general-$p$ one evaluated at $p^*=2$.

We now focus on the main case of this corollary, a slowly closing gap, $\delta = \omega({1}/{\sqrt{L}})$. Then, it is possible to satisfy Eq.~\eqref{pconstraint} for nonzero $p^*$. 
We evaluate the light cone at the optimal value from Eq.~\eqref{pconstraint}, $p^*\approx (L-1)\delta^2$.
The norm simplifies as
\begin{align}
\hspace{-0.7 cm}
\unnorm{\begin{pmatrix}
r \\
1-r
\end{pmatrix}}_\frac{p^*}{p^*-1} &= \left[ r^\frac{p^*}{p^*-1} +(1-r)^\frac{p^*}{p^*-1} \right]^\frac{p^*-1}{p^*}\\
&= \frac{1}{2} \left[ (1+\delta)^\frac{p^*}{p^*-1} +(1-\delta)^\frac{p^*}{p^*-1} \right]^\frac{p^*-1}{p^*}\\
&\hspace{-2.8cm}\approx \frac{1}{2} \left[ 1+ \frac{p^*}{p^*-1}\delta + \frac{p^*}{2(p^*-1)^2} \delta^2 + 1- \frac{p^*}{p^*-1}\delta + \frac{p^*}{2(p^*-1)^2} \delta^2 \right]^\frac{p^*-1}{p^*}\\
&=2^\frac{-1}{p^*}\left[ 1+ \frac{p^*}{2(p^*-1)^2} \delta^2\right]^\frac{p^*-1}{p^*}\\
&\approx 2^\frac{-1}{p^*} \left[ 1+ \frac{1}{2(L-1)}\right].
\end{align}
The overall runtime constraint is then
\begin{align}
    t &\gtrsim (1 - 2\epsilon){R(L)} \frac{2}{\sqrt{p^*}} 2^{(-L+1)/p^*} \frac{1}{\unnorm{\bm \lambda }_{\frac{p^*}{p^*-1}}} \\
    &\approx (1 - 2\epsilon) {R(L)} \frac{2}{\delta \sqrt{L-1}} 2^\frac{-L+1}{p^*} 2^\frac{L-1}{p^*} \left[ 1+ \frac{1}{2(L-1)}\right]^{1-L}\\
    &\approx (1 - 2\epsilon) {R(L)} \frac{2}{\delta \sqrt{L}} \frac{1}{\sqrt{e}}\\
    &\approx (1 - 2\epsilon) \frac{R(L)}{\delta \sqrt{L}}.
\end{align}

\end{document}


\onecolumngrid
\newpage

\setcounter{secnumdepth}{1}
\setcounter{section}{0}
\renewcommand{\thesection}{S\arabic{section}}
\setcounter{thm}{0}
\renewcommand{\thethm}{S\arabic{thm}}
\renewcommand{\thecor}{S\arabic{cor}}
\setcounter{lemma}{0}
\renewcommand{\thelemma}{S\arabic{lemma}}
\setcounter{equation}{0}
\renewcommand{\theequation}{S\arabic{equation}}
\setcounter{table}{0}
\renewcommand{\thetable}{S\arabic{table}}
\setcounter{figure}{0}
\renewcommand{\thefigure}{S\arabic{figure}}
\setcounter{defi}{0}
\renewcommand{\thedefi}{S\arabic{defi}}

\renewcommand{\bibnumfmt}[1]{[S#1]}
\renewcommand{\citenumfont}[1]{S#1}

\newtheorem*{thm1}{Theorem 1}

\title{Supplemental Material for ``Thermal Throttling of Quantum State Transfer''}
\maketitle
 \raggedright
\setlength{\parindent}{20pt}
 \justifying

In this Supplemental Material, we prove and discuss several of the statements in the main text as well as some additional lemmas. In Sec.~\ref{secprops}, we establish some additional properties of approximate state transfer. In Sec.~\ref{seccor1}, we present a proof of a weak converse to Theorem~1 and a proof of Corollary~1. In Sec.~\ref{secmix}, we prove and discuss Theorem~2 and Corollary~2. In Sec.~\ref{secsat}, we provide the remaining saturating protocols for our bounds. In Sec.~\ref{secthm3} we prove Theorem~3. In Sec.~\ref{secp}, we derive the runtime bounds from the operator and Frobenius-norm light cones for a class of thermal states.

\section{Properties of approximate state transfer}\label{secprops}
In this section, we prove some basic properties of approximate state transfer and contrast it with exact state transfer.
Our goal is to understand the dependence of state transfer on the ancilla state. To this end, we define the sets of all ancilla states for which a given protocol $\epsilon$-approximately works,
\begin{equation}
    T_\epsilon\coloneqq \{\sigma: \err(\sigma)\leq\epsilon\}. 
\end{equation}
\begin{lemma}
    $T_\epsilon$ is convex for all $\epsilon \in [0,1]$.
\end{lemma}
\begin{proof}
    For arbitrary $\epsilon$, suppose $\sigma_1,\sigma_2 \in T_\epsilon$. Then, 
    \begin{align}
    \err(\lambda \sigma_1 + (1-\lambda)\sigma_2) &= \tfrac{1}{2}\unnorm{\Lambda_{\lambda \sigma_1 + (1-\lambda)\sigma_2}-\mathds{1}}_\diamond\\
    &=\tfrac{1}{2}\unnorm{\lambda \Lambda_{ \sigma_1} + (1-\lambda)\Lambda_{\sigma_2}-\mathds{1}}_\diamond\\
    &\leq \tfrac{1}{2}\lambda \unnorm{ \Lambda_{ \sigma_1} -\mathds{1}}_\diamond
    + \tfrac{1}{2}(1-\lambda)\unnorm{ \Lambda_{\sigma_2} -\mathds{1}}_\diamond\\
    &=\lambda \err(\sigma_1)+(1-\lambda) \err(\sigma_2)\\
    &\leq\epsilon,
    \end{align}
    where in the second line we used the linearity of $\Lambda$ in $\sigma$ and in the third line the triangle inequality. Thus, any convex combination of $\sigma_1,\sigma_2$ is also in $T_\epsilon$, completing the proof.
\end{proof}

The following lemma restates a result for exact state transfer in this new notation.
\begin{lemma}
    If $\sigma=\sum \lambda_i \dyad{\Phi_i} \in T_0$ with $\lambda_i\neq0$, then all $\dyad{\Phi_i} \in T_0$.
\end{lemma}
\begin{proof}
    See Sec. S1 of the Supplemental Material of \cite{Upadhyaya2026robust}.
\end{proof}
Exact state transfer can be analyzed just via pure ancilla states, since they are the extreme points of $T_0$. In particular, the dimension of the subspace spanned by these pure states quantifies the robustness of the protocol.
However, for nonzero $\epsilon$, the extreme points of $T_{\epsilon>0}$ are not always pure states. It is thus harder to characterize these sets.
However, we can show an inclusion.
\begin{lemma}\label{tracedisterror}
    For an ancilla state $\sigma'$, if there exists $\sigma \in T_\delta$ such that $\frac{1}{2}\unnorm{\sigma'-\sigma}_1<\epsilon$, then $\sigma'\in T_{\delta+\epsilon}$.
\end{lemma}
\begin{proof}
    Let $\sigma'-\sigma=2\Delta$. Apply the primed channel on a dilated space for the diamond norm,
    \begin{align}
    (\Lambda_{\sigma'} \otimes \mathds{1}) (\rho_{ii'}) &= \Tr_\ap(\Xi(\rho_{ii'}\otimes \sigma'_\an) )\\
    &= \Tr_\ap(\Xi(\rho_{ii'}\otimes (\sigma+2\Delta)) )\\
    &= (\Lambda_{\sigma} \otimes \mathds{1}) (\rho_{ii'}) + \Tr_\ap(\Xi(\rho_{ii'}\otimes 2\Delta) ).
    \end{align}
    The trace norm is multiplicative over tensor products and decreasing under channels~\cite{Watrous2018}, so 
    \begin{align}
        \unnorm{\Tr_\ap(\Xi(\rho_{ii'}\otimes 2\Delta) )}_1&\leq \unnorm{\Xi(\rho_{ii'}\otimes 2\Delta) }_1\\
        &\leq\unnorm{(\rho_{ii'}\otimes 2\Delta) }_1\\
        &=\unnorm{\rho_{ii'} }_1\unnorm{2\Delta}_1\\
        &=2\unnorm{\Delta}_1\\
        &<2\epsilon.
    \end{align}
    Therefore, $\tfrac{1}{2}\unnorm{\Lambda_{\sigma'}-\Lambda_{\sigma}}_\diamond<\epsilon$. So by the triangle inequality, $\err(\sigma')<\epsilon+\delta$. Therefore by definition $\sigma'\in T_{\epsilon+\delta}$.
\end{proof}

\section{Results for pure ancilla states}\label{seccor1}

In this section, we present a proof of a weak converse to Theorem~1 and a proof of  Corollary~1.

We start with a weak converse to Theorem~1. 
Assume a channel $\Xi$ satisfies, for all $\ket{\psi}$ and all $\eta = x, y, z$,
\begin{align}
        \abs{{\bra{\psi}_i \bra{\Phi}_\an  S_\eta \ket{\psi}_i\ket{\Phi}_\an} -1}&\leq2\epsilon.
\end{align}
Let $A$ be a single-site operator with $\unnorm{A}_\infty\leq1$ and Pauli decomposition $A=a_0 \id + \sum_\eta a_\eta \sigma_\eta$. For a possibly non-Hermitian operator $\hat{O}$, we have that 
\begin{align}
    \unnorm{\hat O}_\infty &= \unnorm{\Re \hat O + i \Im \hat O}_\infty\\
    &\leq \unnorm{\Re \hat O}_\infty + \unnorm{ \Im \hat O}_\infty\\
    &= \max_{\ket{\psi_1}} \abs {\expval{\Re \hat{O}}{\psi_1}} + \max_{\ket{\psi_2}} \abs {\expval{\Im \hat{O}}{\psi_2}}\\
    &\leq 2 \max_{\ket{\psi}} \abs {\expval{ \hat{O}}{\psi}}. 
\end{align}
So by assumption and Eq.~(31), we have that
\begin{align}
    \unnorm{\sigma_{\eta,i} \Lambda^\dagger_{\dyad{\Phi}}(\sigma_{\eta,f}) -\id }_\infty \leq 4\epsilon.
\end{align}
So,
\begin{align}
\unnorm{\Lambda^\dagger_{\dyad{\Phi}}(A) -A }_\infty&=\unnorm{ \sum_\eta a_\eta \left[\Lambda^\dagger_{\dyad{\Phi}}(\sigma_\eta) - \sigma_\eta\right] }_\infty\\
&\leq \sum_\eta \abs{a_\eta} \unnorm{ \Lambda^\dagger_{\dyad{\Phi}}(\sigma_\eta) - \sigma_\eta }_\infty\\
&= \sum_\eta \abs{a_\eta} \unnorm{ \sigma_\eta\Lambda^\dagger_{\dyad{\Phi}}(\sigma_\eta) - \id }_\infty\\
&\leq 4\epsilon \sum_\eta \abs{a_\eta}.
\end{align}
Further,
\begin{align}
    \sum_\eta \abs{a_\eta} &\leq \sqrt{3} \left( \sum_\eta  \abs{a_\eta}^2 \right)^\frac{1}{2} \\
    &\leq \sqrt{\frac{3}{2}} \unnorm{A}_2\\
    &\leq \sqrt{3} \unnorm{A}_\infty,
\end{align}
using relationships between $p$-norms and properties of the Pauli decomposition.
Thus, $\unnorm{\Lambda^\dagger_{\dyad{\Phi}}(A) -A }_\infty\leq 4 \sqrt{3} \epsilon $. Since this holds for any contraction $A$, $\unnorm{\Lambda^\dagger_{\dyad{\Phi}} -\mathds{1} }_{\infty\rightarrow \infty}\leq 4 \sqrt{3} \epsilon $. As in the proof of Theorem 1, we use the duality of induced norms between maps and their adjoints to conclude 
$\unnorm{\Lambda_{\dyad{\Phi}} -\mathds{1} }_{1\rightarrow 1}\leq 4 \sqrt{3} \epsilon $. Since the channels under consideration are qubit channels, the diamond norm is at most twice the induced trace norm, so we have that 
\begin{equation}
    \frac{1}{2} \unnorm{\Lambda_{\dyad{\Phi}} - \mathds{1} }_{\diamond}\leq 4 \sqrt{3} \epsilon .
\end{equation}

\subsection{Proof of Corollary~1}
In this subsection we prove Corollary~1.
\begin{proof}
A simple calculation shows that $[X_i S_x, Z_i]=-2i\Re V$ with $V\equiv Y_i S_x$~\cite{Upadhyaya2026robust}. 
Observe that 
\begin{align}
\expval{\Re V}  {R,\Phi_k}&=\frac{1}{2}\expval{Y_i S_x + S_x^\dagger Y_i}  {R,\Phi_k} \\
&= \expval{\Re S_x}  {R,\Phi_k}
\end{align}
and similarly with the opposite sign for $\ket{L}$.
For $p<\infty$,
\begin{align}
        2 \norm{\Re V}_p &= 2\cdot 2^{-L/p} \Tr ( \abs{\Re V}^p)^{1/p}\\
        &= 2\cdot 2^{-L/p} \left(\sum_k \expval{\abs{\Re V}^p}  {R,\Phi_k} + \expval{\abs{\Re V}^p}  {L,\Phi_k} \right)^{1/p}\\
        &\geq 2\cdot 2^{-L/p} \left(\sum_k \abs{\expval{\Re V}  {R,\Phi_k}}^p + \abs{\expval{\Re V}  {L,\Phi_k}}^p \right)^{1/p}\\
        &= 2\cdot 2^{-L/p} \left(\sum_k \abs{\expval{\Re S_x}  {R,\Phi_k}}^p + \abs{\expval{\Re S_x}  {L,\Phi_k}}^p \right)^{1/p}\\
        &\geq 2\cdot 2^{-L/p} \left(\sum_k 2\max\{1-2\epsilon_k,0\}^p \right)^{1/p}\\
        &=2\cdot 2^{(-L+1)/p}  \unnorm{\bm w}_{ p},
\end{align}    
where the second line follows from taking the trace in the given basis, the third line follows because $f(x)=\abs{x}^p$ is convex, the fourth line follows from the discussion above, and the fifth line follows from Theorem~1.

At $p=\infty$, 
\begin{align}
\norm{\Re V}_\infty &\geq \max_{\ket{\Phi}} \ \abs{\expval{\Re V}  {R,\Phi}}\\
&\geq \max \bm w.
\end{align}

\end{proof}

\section{Results for Mixed Ancilla States}\label{secmix}
In this section we prove and discuss Theorem~2 and Corollary~2.
We first establish a lemma solving the convex minimization that will later arise in our theorem's proof.
Here, $y_k$ should be understood as $1-2\epsilon_k$ with $\epsilon_k$ the error on the basis state $\ket{\Phi_k}$. 
Evidently, if $\lambda_k=0$, the worst case is that $y_k=0$. Consider from now on only the indices where $\lambda_k>0$.
\begin{lemma}\label{convexminlemma}
The solution to 
     \begin{IEEEeqnarray*}{uC'L}
    $\displaystyle{\argmin_{y_k}}$ & &  \unnorm{\bm y}_p\\
    \yesnumber subject to:& 	& \sum_k \lambda_k y_k  \geq 1-2\epsilon\\
    & &0\leq y_k \leq 1  
\end{IEEEeqnarray*}
is
\begin{equation}
    y_k = \min \{1, \lambda' \lambda_k^{\frac{1}{p-1}}\},
\end{equation}
where the Lagrange multiplier $\lambda'$ is determined by
\begin{equation}
    \sum_k \lambda_k \min \{1, {\lambda'} \lambda_k ^{\frac{1}{p-1}}\} = 1-2\epsilon.
\end{equation}
\end{lemma}

\begin{proof}
WLOG, we can take the first constraint to be an equality. We first solve the extreme cases of $p=1,\infty$. We must have at least one $y_k\geq 1-2\epsilon$ to satisfy the constraint. When $p=\infty$, choosing all $y_k=1-2\epsilon$ is optimal. For $p=1$, we seek to minimize $\sum y_k$ subject to the constraint. The optimal solution is to simply assign maximal weight to the $y_k$ corresponding to the largest eigenvalues.

To solve this optimization in general, we invoke the Karush-Kuhn-Tucker (KKT) conditions, which allow for solving by Lagrange multipliers with inequality constraints~\cite{Boyd2004}. Since the problem is a convex minimization, if a point satisfies the KKT conditions it is not only a local optimum but a global one~\cite{Boyd2004}. 

First, define the Lagrangian, omitting the irrelevant $0\leq y_k$ constraint:
\begin{equation}
    \mathcal{L} = \sum_k y_k^p + \sum_k \mu_k (y_k-1) - \lambda \left(\sum_k \lambda_k y_k -1 +2\epsilon\right)
\end{equation}
with $\mu_k,\lambda\geq0$.
By the first KKT condition, we desire a stationary point of $\mathcal{L}$. Compute 
\begin{equation}
\frac{d}{d y_j} \mathcal{L} = p y_j^{p-1} + \mu_j - \lambda \lambda_j.
\end{equation}
The second KKT condition of complementary slackness is that $\mu_j (y_j-1)=0$ for all $j$. If $\mu_j=0$, which satisfies the third KKT condition of dual feasibility, then by stationarity $ y_j=  \left(\frac{\lambda}{p} \lambda_j \right)^{\frac{1}{p-1}}$. By the fourth KKT condition of primal feasibility, if this value of $y_j$ is $\leq 1$, we are done. Else, $\mu_j$ must not equal $0$ so $y_j=1$ by complementary slackness. 
Define a rescaled Lagrange multiplier $\lambda'=(\lambda/p)^{\frac{1}{p-1}}$. We can summarize this state of affairs by writing
\begin{equation}
    y_j = \min \{1, \lambda' \lambda_j^{\frac{1}{p-1}}\},
\end{equation}
where the Lagrange multiplier is determined by the condition 
\begin{equation}
    \sum_k \lambda_k \min \{1, {\lambda'} \lambda_k ^{\frac{1}{p-1}}\} = 1-2\epsilon.
\end{equation}
In the limit $p\rightarrow \infty$, $\lambda'\rightarrow 1-2\epsilon$, which agrees with our earlier analysis.

\end{proof}

Solving for $\lambda'$ corresponds to solving the dual to the original problem. A feasible lower bound on $\lambda'$ gives a lower bound to the original problem. Observe that the constraint function is increasing with $\lambda'$, and that it suffices to consider the range $[0,{\lambda_\text{min}^{1/(1-p)}}] $. Organize the eigenvalues with degeneracies $D_j$ and unique values in decreasing order $v_j$. The constraint can then be expressed as 
\begin{equation}
    1-2\epsilon= \sum_{j<c} D_j v_j + \sum_{j\geq c} D_j \lambda' v_j^{\frac{p}{p-1}},
\end{equation}
where $v_c$, if it exists, is the largest eigenvalue such that $\lambda' v_{c}^\frac{1}{p-1}<1$. A smaller $\lambda'$ leads to a smaller $c$.
This gives a $p$-independent upper bound on $c$ and in turn $\lambda'$---the sum $\sum_{j<c} D_j v_j$ cannot exceed $1-2\epsilon$.

\subsection{Proof of Theorem~2}
We now prove Theorem~2.
\begin{proof}
    Let $\sigma= \sum_k \lambda_k \dyad{\Phi_k} $. As usual, $ \norm{[X_i S_x, Z_i]}_p = 2\cdot 2^{-L/p} \unnorm{\Re V}_p$.
    Key steps from Theorem~1 extend to mixed states:
    By the argument leading to Eq.~(29) and linearity,
    \begin{equation}
    \Tr_\an(S_x (\id_i\otimes \sigma_\an)) = X_i \Lambda^\dagger_{\sigma}(X_f),    \label{adjstab}
    \end{equation}
    and by the argument following Eq.~(32),
    \begin{equation}\label{opnormdiff}
        \unnorm{X_i \Lambda^\dagger_{\sigma}(X_f)-\id_i}_\infty \leq 2\epsilon.
    \end{equation}
    The above equation implies
    \begin{align}
        1-\Re \bra{R} X_i \Lambda^\dagger_{\sigma}(X_f) \ket{R} &\leq 2\epsilon,\\
        1-\Re \bra{L} X_i \Lambda^\dagger_{\sigma}(X_f) \ket{L} &\leq 2\epsilon.
    \end{align}

    We now formulate the lower bound as an optimization in the following variables: 
    \begin{align}
    r_k&\coloneqq \expval{\Re S_x}{R,\Phi_k}=\expval{\Re V}{R, \Phi_k} ,   \\
    l_k&\coloneqq \expval{\Re S_x}{L,\Phi_k}=-\expval{\Re V}{L, \Phi_k}.
    \end{align}
    These variables are constrained as follows:
    \begin{align}
        1-2\epsilon &\leq \Re \bra{R} X_i \Lambda^\dagger_{\sigma}(X_f) \ket{R}\\
        &= \Re \Tr(S_x (\dyad{R}_i \otimes \sigma_\an))\\
        &= \Tr(\Re S_x (\dyad{R}_i \otimes \sigma_\an))\\
        &=\sum_k \lambda_k r_k,
    \end{align}
    and the same for $l_k$.
    On the other hand, we desire a lower bound on 
    \begin{align}
         \unnorm{ \Re V}_p &= \Tr(\abs{\Re V}^p)^{1/p}\\ 
         &= \left(\sum_k \expval{\abs{\Re V}^p}{R, \Phi_k}+\expval{\abs{\Re V}^p}{L, \Phi_k}\right)^{1/p} \\
         &\geq  \left(\sum_k \abs{r_k}^p + \abs{l_k}^p\right)^{1/p}.
    \end{align}   
    Therefore, $\unnorm{ \Re V}_p\geq W$ with
    \begin{IEEEeqnarray*}{uC'L}
    $W=\displaystyle{\minimize_{r_k,l_k}}$ & &\left(\sum_k \abs{r_k}^p+\abs{l_k}^p\right)^{1/p}\\
    \yesnumber subject to:& 	& \sum_k \lambda_k r_k  \geq 1-2\epsilon\\
    & &  \sum_k \lambda_k l_k  \geq 1-2\epsilon\\
    & &-1\leq l_k, r_k \leq 1.  
\end{IEEEeqnarray*}

The $r_k$ and $l_k$ optimizations are independent and identical so we need only solve one. Moreover, negative values of the unknown variables increase the objective function without helping satisfy the constraint so WLOG we can take the variables to be nonnegative. Finally, the inequality constraint being saturated is the worst case, and is always achievable by choosing all variables to be $1-2\epsilon$, so we can take it to be an equality constraint. Altogether,
    \begin{IEEEeqnarray*}{uC'L}\label{thmconvopt}
    $\frac{W}{2^{1/p}}=\displaystyle{\minimize_{y_k}}$ & &  \left(\sum_k y_k^p\right)^\frac{1}{p}\\
    \yesnumber subject to:& 	& \sum_k \lambda_k y_k  = 1-2\epsilon\\
    & &0\leq y_k \leq 1  .
\end{IEEEeqnarray*}
This optimization formulation clarifies why commutator bounds for state transfer over a single mixed state resemble those over the ensemble of pure eigenstates with specially chosen errors.
Invoking Lemma~\ref{convexminlemma} and observing $\norm{ \Re V}_p\geq 2^{-L/p}  W = 2^{(-L+1)/p} \unnorm{\bm y}_p=\norm{\bm y}_p$, we are done.    
\end{proof}

\subsection{Proof of Corollary~2}
\begin{proof}
    Starting from Eq.~\eqref{thmconvopt}, we use H\"{o}lder's inequality:
    \begin{align}
        \unnorm{\bm y}_p &\geq \frac{\bm y \cdot \bm \lambda }{\unnorm{\bm \lambda}_{\frac{p}{p-1}}}\\
        &\geq \frac{1-2\epsilon}{\unnorm{\bm \lambda}_{\frac{p}{p-1}}}.
    \end{align}
    (Where the H\"{o}lder dual of $p=1$ is $p=\infty$.)
\end{proof}

\subsection{Proof of monotonicity with \texorpdfstring{$p$}{}}
This follows for the theorem, since the renormalized $p$-norms are increasing with increasing $p$. Suppose we have an optimal solution $\bm {y}^*$ for a specific $p_1$. Since the constraints don't depend on $p$, this solution continues to be feasible for $p_2<p_1$. But we minimize over $\bm y$ so the actual solution for $p_2$ can only be smaller. For the corollary, we use the well-known relation for $p$-norms, which can be proven using H\"{o}lder's, that 
\begin{equation}
    \unnorm{\bm \lambda}_{\frac{p_1}{p_1-1}} \leq 2^{(L-1)(\frac{p_1-1}{p_1}-\frac{p_2-1}{p_2})} \unnorm{\bm \lambda}_{\frac{p_2}{p_2-1}}.
\end{equation}

\subsection{Comparison of theorem and corollary}
\begin{figure*}[t!]
    \centering
        \includegraphics[height=1.9in]{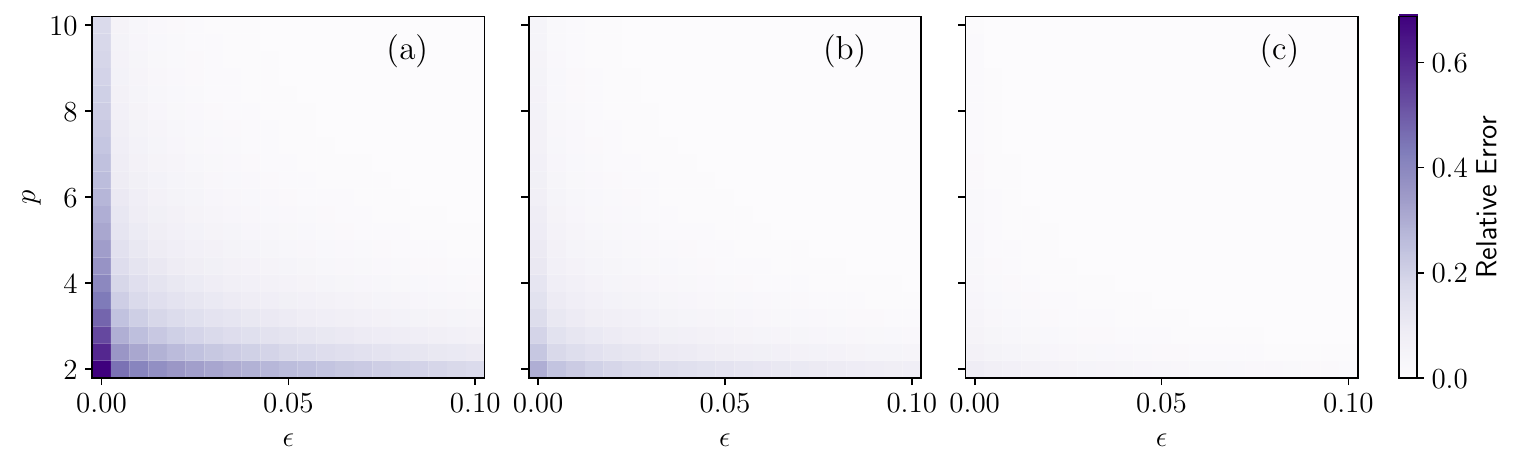}
    \caption{Relative difference between lower bounds from Theorem~2 and Corollary~2 for $L=20$. Average computed over 50 spectra sampled from $\text{Dir}(\alpha')$ with scales (a) $\alpha'=0.1$, (b) $\alpha'=1$, and (c) $\alpha'=5$. The difference is negligible except when $p$ and $\epsilon$ are small and $\bm \lambda$ is far from uniform.}
    \label{bounddiff}
\end{figure*}

We compare the lower bounds from Theorem~2 and Corollary~2 in Fig.~\ref{bounddiff}. With $L=20$, we sample 50 different spectra $\bm \lambda$ from a symmetric Dirichlet distribution specified by a scale $\alpha'$, $\text{Dir}(\alpha')$~\cite{dirichlet}. As $\alpha'$ is increased, the samples bias towards distributions that are closer to maximally mixed, with $\alpha'=1$ corresponding to uniform sampling over the probability simplex. We sweep over $\epsilon$ and $p$ values, at each point numerically solving for the Lagrange multiplier and plotting the average relative error between the resultant theorem bound and the corollary bound. 

The figure supports our claim that the theorem and corollary bounds are close across three axes. They almost agree over large swathes of parameter space. The largest difference is at small $p$ and small $\epsilon$ and decays away from this. This trend stays the same as we consider distributions that are flatter, but the actual value of the difference decreases.

\section{Saturating protocols}\label{secsat}
In this section we present the remaining saturating protocols for our bounds. The protocols are built around dephasing in the $Y$ basis.

\subsection{Corollary 1}
Assume $2\epsilon_k\leq1$ and consider $\Xi$ defined by Kraus operators $\{\mathrm{SWAP}_{i,f}  \sqrt{1-\epsilon_k} \id_i \otimes \dyad{\Phi_k}_\an, \mathrm{SWAP}_{i,f}  \sqrt{\epsilon_k} Y_i \otimes \dyad{\Phi_k}_\an \}_k$. As for Theorem~1's saturating protocol, the effective transfer channels are dephasing in the $Y$ basis, $\Lambda_{\dyad{\Phi_k}}(\rho)=(1-\epsilon_k) \rho+\epsilon_k Y\rho Y$. 
Following the same arguments as above, it can be verified that $\Xi$ achieves $\epsilon_k$-approximate state transfer on $\ket{\Phi_k}$. We compute the commutator using the adjoint Kraus operators,
\begin{align}
    \Xi^\dagger(X_f) &= \sum_k \left[(1-\epsilon_k)  X_i + \epsilon_k Y_i X_i Y_i\right] \otimes \dyad{\Phi_k}\\
     &= \sum_k (1-2\epsilon_k)  X_i \otimes \dyad{\Phi_k}.
\end{align}
Then, 
\begin{align}
    [X_i S_x,Z_i] &= [\Xi^\dagger(X_f),Z_i]\\
    &= -\sum_k (1-2\epsilon_k) 2i Y \otimes \dyad{\Phi_k},
\end{align}
so
\begin{align}
    \norm{[X_i S_x,Z_i]}_p&=2\cdot 2^{-L/p} \unnorm{(1-2 \bm \epsilon,1-2 \bm \epsilon)}_p\\    
    &=2\cdot  2^{(-L+1)/p} \unnorm{1-2 \bm \epsilon}_p\\
    &=2 \norm{\bm w}_p.
\end{align}

We next show that unitary protocols can get within $2^{1/p}$ of the bound. Our strategy is to implement dephasing channels conditioned on half the ancilla states---it is not clear how to implement them on all ancilla states without using extra dilating qubits. Then, we augment the dephasing with local unitary rotations, which guarantee that the protocol achieves the right errors, but lead to a small factor difference in the commutator norm. Again assume $2\epsilon_k\leq1$ and that the errors are ordered from smallest to largest. Similar to above, define effective Pauli operators on paired subspaces, $\tilde{Y}_k = -i \dyad{\Phi_k}{\Phi_{k+2^{L-2}}} +i\dyad{\Phi_{k+2^{L-2}}}{\Phi_k}$. Consider the unitary 
\begin{equation}
    U_\text{deph}= \sum_{k=1}^{2^{L-2}} \left( \sqrt{1-\epsilon_k} \id_i \otimes \tilde{\id}_k + i\sqrt{\epsilon_k} Y_i \otimes \tilde{Y}_k  \right).
\end{equation}
Combined with $\mathrm{SWAP}_{i,f}$, this unitary realizes $Y$-dephasing of strength $\epsilon_k$ on $\ket{\Phi_k}$ and on $\ket{\Phi_{k+2^{L-2}}}$. To match all the errors, we concatenate with single-qubit rotation errors conditioned on $\ket{\Phi_k}$ with $k>2^{L-2}$:
\begin{equation}
    U_\text{rot}= \sum_{k=1}^{2^{L-2}} \id_i \otimes \dyad{\Phi_k}+ \sum_{k=2^{L-2}+1}^{2^{L-1}} e^{-i \theta_k Z_i} \otimes \dyad{\Phi_k}.
\end{equation}
With an appropriate choice of $\theta_k$, $\mathrm{SWAP}_{i,f}  U_\text{deph} U_\text{rot} $ achieves $\epsilon_k$ error on all $\ket{\Phi_k}$. On the other hand, the commutator is 
\begin{equation}
    [U^\dagger_\text{deph} X_i U_\text{deph}, U_\text{rot} Z_i U_\text{rot}^\dagger]= -\sum_{k=1}^{2^{L-2}} (1-2\epsilon_k) 2i Y \otimes (\dyad{\Phi_k}+\dyad{\Phi_{k+2^{L-2}}}),
\end{equation}
since the single-qubit rotations do not affect $Z_i$. Then,
\begin{align}
    \norm{[U^\dagger X_f U,Z_i]}_p&=\norm{ U_\text{rot} [U^\dagger X_f U,Z_i] U_\text{rot}^\dagger }_p\\
    &=\norm{  [U^\dagger_\text{deph} X_i U_\text{deph},U_\text{rot} Z_i U_\text{rot}^\dagger] }_p\\
    &=2\cdot 2^{-L/p} \left(2 \sum_{k=1}^{2^{L-2}} 2(1-2\epsilon_k)^p\right)^{1/p}\\
    &= 2^{1/p} \cdot 2\cdot  2^{(-L+1)/p} \left(\sum_{k=1}^{2^{L-2}} (1-2\epsilon_k)^p\right)^{1/p}\\
    &\leq 2^{1/p} \cdot 2\cdot  2^{(-L+1)/p} \left(\sum_k (1-2\epsilon_k)^p\right)^{1/p}\\
    &= 2^{1/p} \cdot 2 \norm{\bm w}_p,
\end{align}
where the first line follows by the norms' unitary invariance. We note that this construction does not similarly guarantee a $2^{1/p}$ gap for other commutators like $[X_i S_x,Y_i]$ and $[Z_i S_z,X_i]$.

\subsection{Theorem 2}
The saturating construction is similar to the previous one. Instead of the dephasing strengths being specified by $\epsilon_k$, let them be specified by $\frac{1-y_k}{2}$. By the constraint on the Lagrange multiplier, this channel achieves average error $\epsilon$. By the same argument as above, the commutator norm is $2\norm{\bm y}_p$.

\section{Proof of Theorem~3}\label{secthm3}
In this section we prove Theorem~3.
Before proving the theorem, we remark that in principle the strongest bound would be from using Theorem~2 with the $p$-norm light cone, i.e.
\begin{equation}
     t \gtrsim \max_p 2\norm{\bm y(\bm \lambda, p)}_p  \frac{R(L)}{ \sqrt{p} C}, 
\end{equation}
where we have made explicit the dependence of $\bm y$ on other parameters. Because solving for $\bm y$ requires solving for the Lagrange multiplier, it is more challenging to manipulate this optimization analytically. Hence, we focus on the bound from Corollary~2. As we have discussed, the bounds from Corollary~2 and Theorem~2 are close in several regimes.

We present some useful identities for the proof. Let $\bm r$ be a probability vector. Again let $\bm r(q)$ be the probability vector defined by exponentiating elementwise $r_k \rightarrow r_k^q$ then renormalizing. Then,
\begin{equation}
    \frac{d}{dq} \ln \unnorm{\bm r}_q = \frac{-1}{q^2} H(\bm r(q)),
\end{equation}
where $H$ is the Shannon entropy. This identity follows from 
\begin{align}
    \frac{d}{dq} \ \frac{1}{q}\ln \sum_i r_i^q &= \frac{1}{q^2} \left(-\ln \sum_i r_i^q + \frac{\sum_i r_i^q \ln r_i^q }{\sum_i r_i^q} \right).
\end{align}
The second identity relates the rate of change of entropy to the variance of the random variable $\ln \bm r$,
\begin{align}
    \frac{d}{dq} H(\bm r(q)) = -q \text{Var}_{\bm r(q)}[ \ln \bm r],
\end{align}
where $\ln \bm r $ is the elementwise logarithm.
Applying the first identity above gives, after some algebra,
\begin{equation}
    \frac{d}{dp} \ln {\unnorm{\bm \lambda }_{\frac{p}{p-1}}} = \frac{1}{p^2} H\left[\bm \lambda\left(\frac{p}{p-1}\right)\right].
\end{equation}

The objective function tends to zero at large $p$, so is maximized either at $p=2$ or where its derivative is zero. We compute the derivative,
\begin{align}
    \hspace{-0.5cm}\frac{d}{dp} \ln\left[ \frac{2^{(-L+1)/p}}{\sqrt{p}}  \frac{1}{\unnorm{\bm \lambda }_{\frac{p}{p-1}}}\right]&= \frac{d}{dp}\left( \ln \frac{1}{\sqrt{p}} + \ln 2 \frac{-L+1}{p} - \ln {\unnorm{\bm \lambda }_{\frac{p}{p-1}}}\right)\\
    & \hspace{-1cm} =  
    \frac{-1}{2p} - \frac{1}{p^2} H\left[\bm \lambda\left(\frac{p}{p-1}\right)\right]+\frac{L-1}{p^2} \ln 2 \label{slope}.
\end{align}
Thus, the derivative is zero where
\begin{align}\label{saddlept}
\frac{p}{2} + H\left[\bm \lambda\left(\frac{p}{p-1}\right)\right]&=(L-1) \ln 2 .
\end{align}
The entropy $H(\bm r(q))$ decreases with $q$ since $\bm r(q)$ becomes more concentrated on its largest values (this can be proven using the second identity above). Thus, the left-hand side of Eq.~\eqref{saddlept} is strictly increasing with $p$. This means that if a solution to Eq.~\eqref{saddlept} exists, it is unique, and that the equation's left-hand side takes its minimum value at $p^*=2$. This minimum value is $1 + H\left[\bm\lambda\left(2\right)\right]$. If this is greater than $(L-1) \ln 2$, then there is no critical point and the maximum is at $p=2$. Otherwise, there exists a unique critical point $p^*\geq 2$ since the left-hand side is unbounded as a function of $p$. By Eq.~\eqref{slope}, the derivative at $p=2^+$ is positive so the maximum is at this critical point.

\section{Static-\texorpdfstring{$p$}{} bounds for thermal states}\label{secp}
In this section, we derive the runtime lower bounds from Theorem~2 combined with the operator-norm or Frobenius-norm light cones for a specific class of mixed ancilla states.

In Theorem~2, the operator norm can be simply evaluated to give a lower bound of $2(1-2\epsilon)$, which, when plugged into the operator-norm light cone, gives the bounds stated in the main text.
The Frobenius-norm bound calculation is more involved. Here, let $R_F(L)$ denote the Frobenius light cone, analogous to the $R(L)$ light cone for general $p$-norms mentioned in the main text. As a warm-up, consider the bound from the Frobenius light cone combined with Corollary~2:
\begin{align}
    t \gtrsim R_F(L) 2^{(-L+1)/2} \frac{1}{\unnorm{\bm \lambda }_{2}}.
\end{align}
Now, we can compute 
\begin{align}
    {\unnorm{\bm \lambda }_2} &= {\unnorm{(r,1-r)}^{L-1}_{2} }\\
    &\approx \left[ \frac{1}{\sqrt{2}} \left(1+\frac{1}{2L^\gamma}\right) \right]^{L-1},
\end{align}
so that 
\begin{align}
    t &\gtrsim R_F(L) \left(1+\frac{1}{2L^\gamma}\right) ^{-L}\\
      &\approx R_F(L) e^{-\frac{1}{2}L^{1-\gamma}} \\
      &\rightarrow 0,
\end{align}
where the second line follows from $e=\lim_{n\rightarrow \infty} (1+1/n)^n$ and the last since $\gamma<1$.

We next show that even the possibly stronger theorem bound is trivial. For readability, we suppose that the ancilla state consists of $L$, not $L-1$, qubits. For the thermal state we consider, the eigenvalues are $r^j (1-r)^{L-j}$ with degeneracies $\binom{L}{j}$. The condition on the Lagrange multiplier is then
\begin{equation}\label{p2constraint}
    1-2\epsilon = \sum_{j=c+1}^{L} \binom{L}{j} r^j (1-r)^{L-j} + \lambda' \sum_{j=0}^{c} \binom{L}{j} r^{2j} (1-r)^{2(L-j)},
\end{equation}
where the cutoff $c$ satisfies 
\begin{equation}\label{lagrangeconstraint}
 \lambda' r^{c} (1-r)^{L-c}<1.   
\end{equation}
The solution vector can be decomposed based on whether the minimum condition in Eq.~(10) 
is satisfied: $\bm y=\bm y_1 + \bm y_2 $ where ${y_2}_k= y_k $ if $\lambda' \lambda_k \leq 1$ and ${y_2}_k=0$ otherwise. By the triangle inequality, $\norm{\bm y}_2 \leq \norm{\bm y_1}_2 + \norm{\bm y_2}_2 $. We show both terms go to zero even when multiplied by $R_F(L)$---$R_F(L)$ is at most linear in $L$~\cite{Tran_2020_hierarchy,Chen2021}. 

To do so, we derive a lower bound on $c$. Eq.~\eqref{p2constraint} implies that $\sum_{j=c+1}^{L} \binom{L}{j} r^j (1-r)^{L-j}\leq 1-2\epsilon$, or equivalently, $\sum_{j=0}^{c} \binom{L}{j} r^j (1-r)^{L-j}\geq 2\epsilon$. By the additive Chernoff bound on the Bernoulli distribution of $L$ tosses of an $r$-weighted coin, this condition is violated if
\begin{equation}\label{lbc}
    c<Lr - \sqrt{\frac{L}{2} \ln \frac{1}{2\epsilon}}=Lr - A\sqrt{L}.
\end{equation}
The first term can be bounded again by the Chernoff bound, now for $L$ tosses of a fair coin:
\begin{align}
    \norm{\bm y_1}_2^2 &= 2^{-L} \sum_{j=c+1}^{L} \binom{L}{j}\\
    &\leq \exp{-2\left(\frac{L\delta}{2} - A \sqrt{L}\right)^2/L}\\
    &= \exp{-\frac{L\delta^2}{2} + 2A \delta\sqrt{L} - 2A^2}\\
    &= \exp{-\frac{L^{1-\gamma}}{2} + 2A L^\frac{1-\gamma}{2} - 2A^2}\\
    &\approx \exp{-\frac{L^{1-\gamma}}{2}}.
\end{align}
Thus, $2R_F(L)\norm{\bm y}_1\rightarrow 0$.
The second term is bounded by Eq.~\eqref{p2constraint}:
\begin{align}
    \norm{\bm y_2}_2^2 &= \lambda'^2 2^{-L} \sum_{j=0}^{c} \binom{L}{j} r^{2j} (1-r)^{2(L-j)}\\
    &\leq \lambda' 2^{-L} (1-2\epsilon).
\end{align}
We need only show $\lambda' 2^{-L}\rightarrow 0$, which follows from Eqs.~\eqref{lagrangeconstraint}, \eqref{lbc}:
\begin{align}
    \lambda' 2^{-L}&\leq (1+\delta)^{-c} (1-\delta)^{c-L}\\
    &\approx  \exp{-\frac{L\delta}{2} -\frac{L\delta^2}{4}} \exp{\frac{L\delta}{2}- \frac{L\delta^2}{4}}\\
    &\approx  \exp{-\frac{L\delta^2}{2}}.
\end{align}
Thus, $2R_F(L)\norm{\bm y}_2\rightarrow 0$.

\bibliography{mainbib}